\RequirePackage{fix-cm}

\documentclass[smallextended]{svjour3}       % onecolumn (second format)

\smartqed  % flush right qed marks, e.g. at end of proof

\usepackage{mathptmx}      % use Times fonts if available on your TeX system12
\usepackage{graphicx}
\usepackage{amsmath}
\usepackage{amssymb}

\journalname{Experimental Astronomy}

\begin{document}

\title{Bandpass calibration of a wideband spectrometer using coherent pulse injection}
\titlerunning{Wideband spectrometer calibration}        

\author{Nipanjana Patra \and Justin D. Bray \and Paul Roberts \and Ron D. Ekers }
\authorrunning{N.\ Patra et al.} 

\institute{
 Nipanjana Patra
 \at Raman Research Institute, C V Raman Avenue, Sadashivanagar, Bangalore 560080, India \\ \email{nipanjana@rri.res.in}\\
  \emph{Current address for correspondence :} University of California, Berkeley, CA 94720, US \\
  \email{nipanjana@berkeley.edu}
\and
 Justin D. Bray
  \at JBCA, School of Physics \& Astronomy, University of Manchester, Manchester M13 9PL, UK
\and
 Paul Roberts \and Ron D. Ekers
  \at CSIRO Astronomy \& Space Science, Epping, NSW 1710, Australia
}

\maketitle

\begin{abstract}
We present a relatively simple time domain method for determining the bandpass response of a system by injecting a nanosecond pulse and capturing the system voltage output.  A pulse of sub-nanosecond duration contains all frequency components with nearly constant amplitude up to 1~GHz.  Hence, this method can accurately determine the system bandpass response to a broadband signal. In a novel variation on this impulse response method,  a train of pulses is coherently accumulated providing precision calibration with a simple system. The basic concept is demonstrated using a pulse generator-accumulator setup realised in a Bedlam board which is a high speed digital signal processing unit. The same system was used at the Parkes radio telescope between 2--13 October 2013 and we demonstrate its powerful diagnostic capability. We also present some initial test data from this experiment.

\end{abstract}


\section{Introduction}
\label{sec:intro} 
Precise calibration of the bandpass of a radio telescope is critical for a number of astronomical observations, including searches for the spectral signature of the cosmological evolution of neutral hydrogen~\cite{Shaver99,Patra13} and for baryon acoustic oscillations~\cite{Aubourg}.  The bandpass of an instrument is most commonly calibrated in the frequency domain, by measuring the spectrum of an injected calibration signal, but it can also be calibrated in the time domain, by injecting a time-varying signal, such as a pulse, and recording the pulse profile after it has passed through the instrument.
The recorded pulse profile provides an indication of the impulse response of the instrument, and its Fourier transform --- divided by that of the original pulse --- represents the complex spectral response.

In this work, we extend this impulse response technique by generating a regular series of pulses with timing locked to our sampling rate, allowing us to coherently sum the profiles of the individual pulses.
Compared to using individual pulses or calibration with an incoherent broadband noise source, this allows us to rapidly extract an accurate measure of the impulse response of the instrument, while keeping the amplitude of individual pulses sufficiently low to avoid non-linear behaviour. 
The required processing is minimal compared to frequency-domain methods utilising spectrometers, which must calculate the spectrum in real time. Implementation is also simpler than for a similar technique used in telecommunications in which a known pseudo-random sequence is recorded instead. The generated pulses are well-measured and substantially shorter than the inverse bandwidth of our system, constituting a well-understood flat-spectrum calibration signal. The total power in individual pulses is negligible relative to the system noise, so that the pulse injection system could be run continuously without degrading the system performance. Our approach constitutes a simple technique for accurately calibrating the complex spectral response of a radio telescope.

In section~\ref{sec:experiment}, we describe the hardware used in the implementation of our technique, consisting of a fast pulse generator and a Bedlam board~\cite{Bray13} used for digital signal-processing.  In section~\ref{sec:lab} we describe a benchtop implementation of our technique in a system similar to the SARAS spectrometer~\cite{Patra13}, in which the signal propagation paths and other non-idealities are well understood, to demonstrate its functionality.  In section~\ref{sec:parkes}, we show some results from the implementation of our technique on an active astronomical instrument, the Parkes radio telescope.  Finally, we conclude briefly in section~\ref{sec:conc}.

\section{System description}
\label{sec:experiment}

\begin{figure}[!htb]
\centering
\includegraphics[ angle=0, width=10cm]{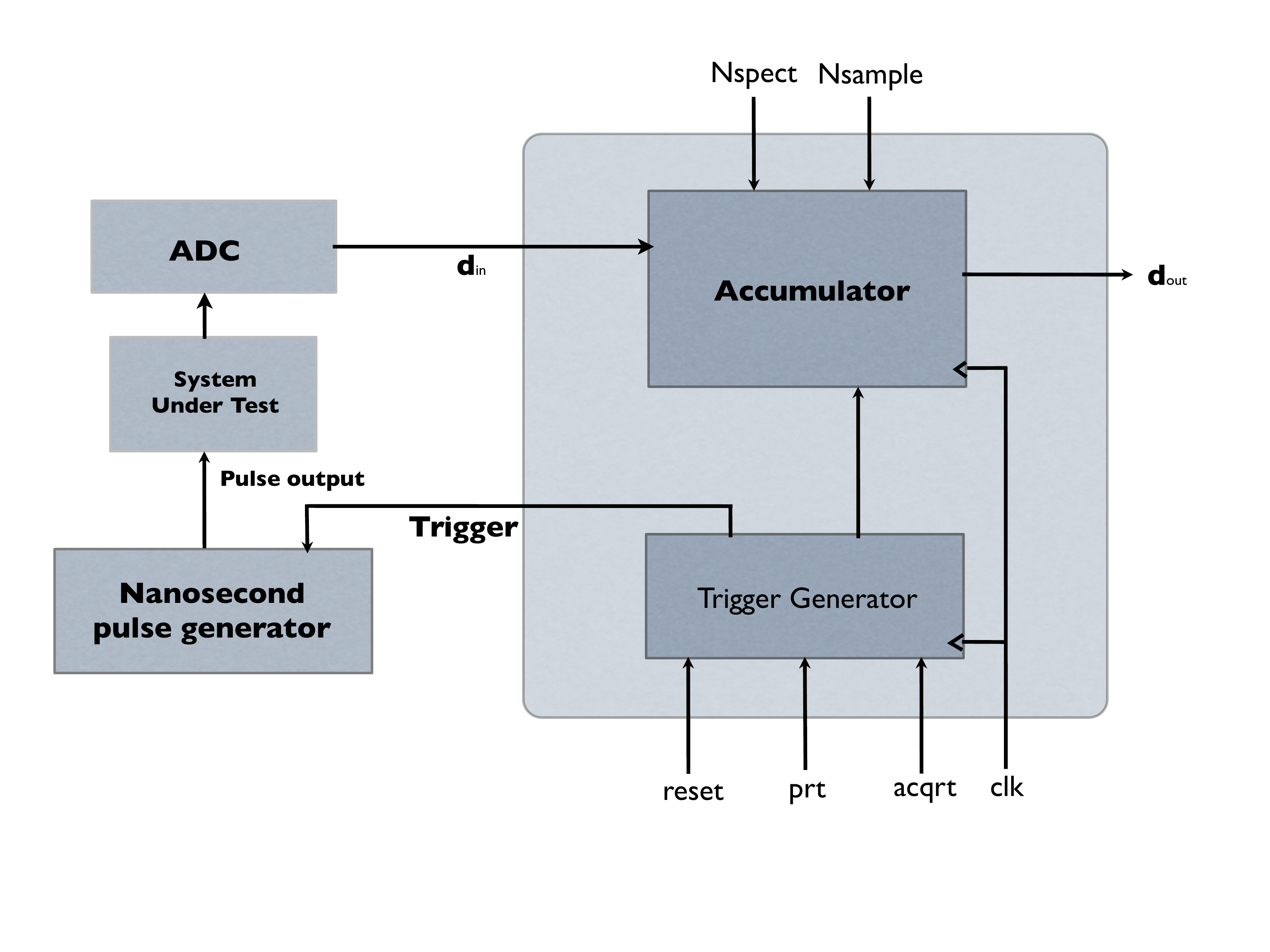}
\caption{Schematic of the pulse accumulator setup.}
\label{Accum}
\end{figure}

The basic experimental setup for system bandpass calibration is shown in Fig.~\ref{Accum}. We inject a nanosecond pulse to a system being calibrated. System responses to several pulses are coherently added in an accumulator.  The system components are described below.

\subsection{Nanosecond pulse generator}
\label{sec:pulsegen}

\begin{figure}[ht]
\begin{minipage}[b]{0.5\linewidth}
\centering
\includegraphics[angle=0,  width =\linewidth]{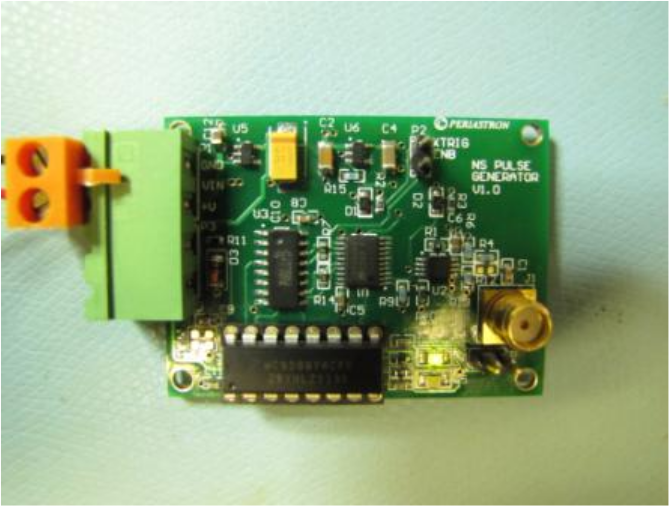}
%\caption{}
%\label{pulse_gen1}
\end{minipage}
\begin{minipage}[b]{0.5\linewidth}
\centering
\includegraphics[angle=0, width =\linewidth]{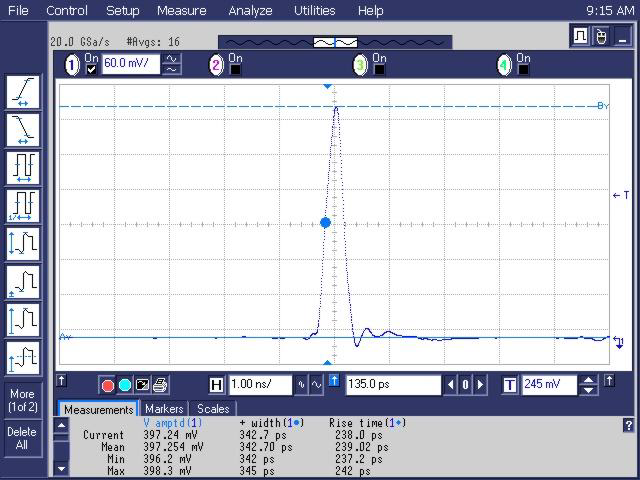}
\end{minipage}
\caption{Nanosecond pulse generator (Left). A typical output pulse voltage seen in an oscilloscope (Right). The horizontal ticks are 0.1~ns. Sampling rate of the oscilloscope is 20GS/sec and bandwidth is 6 GHz.}
\label{pulse_gen2}
\end{figure}

The pulse generator we used for this work and the output pulse it produces are shown in Fig.~\ref{pulse_gen2}. This pulse is injected into the system being measured and the system output is sampled.

Since radio telescope receivers have high gain and sensitivity only a low voltage pulse is required, as it is desirable to keep the input voltage excursion of the pulse below the input noise power level, to ensure the system retains a linear response to the pulse input. The pulse is generated with a custom-built circuit based on a high speed positive emitter coupled logic (PECL) D-type flip-flop with reset. The pulse generator board works in two modes: Asynchronous triggering from an on board time base or externally triggered from a supplied event. The generated pulse has the following properties:

\begin{itemize}
\item
Output pulse width : 350~ps (50\% width).
\item
Output rise/fall time : 240~ps.
\item
Pulse amplitude : 400~mV unipolar pulse into 50~$\Omega$.
\end{itemize} 
The low intrinsic jitter of 0.2~ps for the PECL D-type flip-flop allows synchronous integration of the received pulses to be performed to high precision as this is negligible compared to the sample period of the following Analog to Digital Converter (ADC), which is approximately 1~ns. The narrow pulse width and rise/fall time ensure the pulse contains spectral content up to a few GHz, which was sufficient for our purposes. The power requirement for the pulse board is 3.6~V at 100~mA. 

%We note that the injected pulse is not a perfect delta function but has finite width and some asymmetry. If it is necessary to obtain the absolute phase and the amplitude calibration we would have to measure and deconvolve these intrinsic pulse properties.
As our injected pulse is not a perfect delta function, having non-zero width and some asymmetry, strictly it is necessary to deconvolve the injected pulse from the received pulse profile in order to achieve absolute phase and amplitude calibration.  However, as the pulse width is substantially narrower than the inverse of the 300--350~MHz system bandwidth in sections~\ref{sec:lab} and~\ref{sec:parkes}, it is close to a perfect delta function at these timescales, and so we omit this step.

\subsection{Bedlam Board}

In our experiments, the trigger signal is sourced from a custom general purpose high speed digital signal processing board, known as Bedlam. The Bedlam board consists of 8 RF input channels feeding two 8-bit quad-channel ADCs (EV8AQ160). The digital sample data from each of two channels of the ADCs are directed to one of four DSP oriented FPGAs (XC5VSX95T). A final logic oriented FPGA (XC5VLX30T) provides dual fast Ethernet and general purpose I/O interface from the board.
 
\begin{figure}[ht]
\center
\includegraphics[angle=0, width=8.0cm]{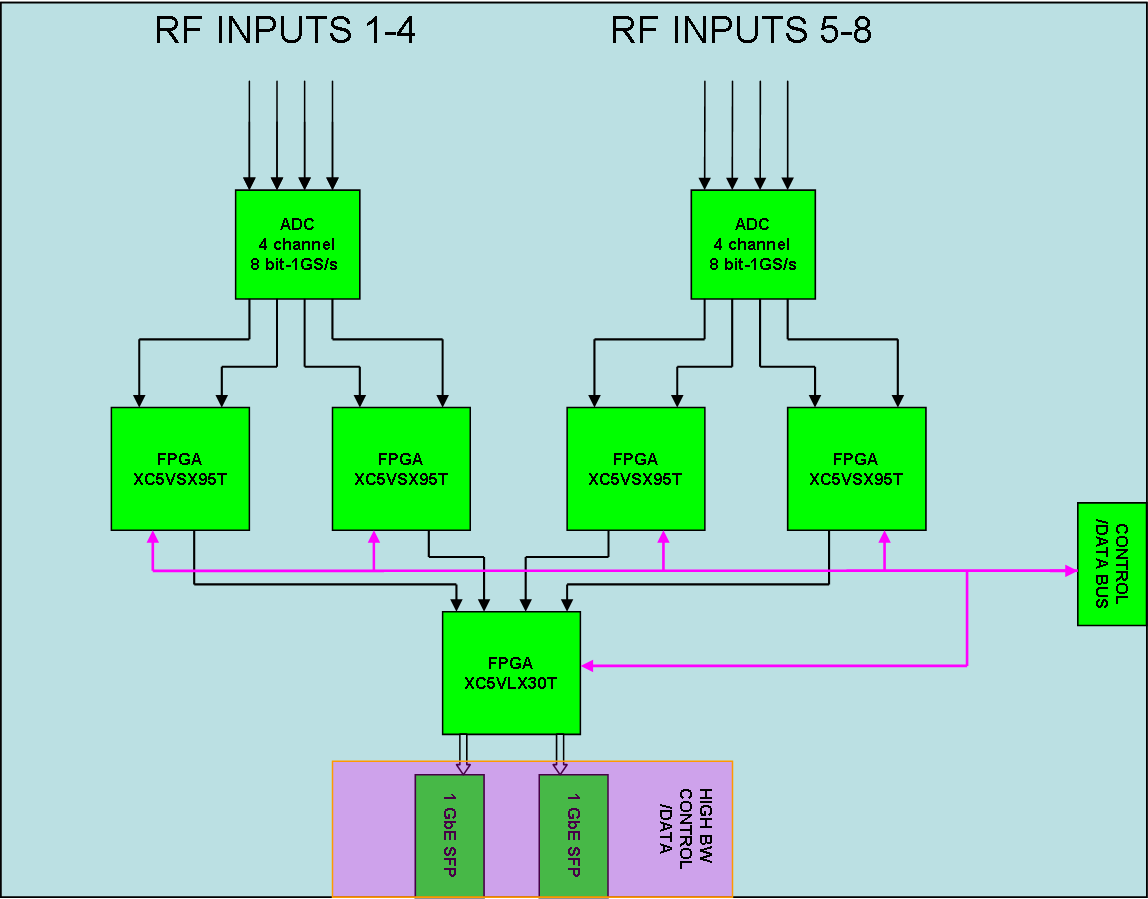}
\caption{Bedlam board schematic.}
\label{Bedlam_schem}
\end{figure}

The pulse generator output is fed into the system being calibrated and the system output is captured by the Bedlam board (Fig.~\ref{Accum}). In order to improve the signal-to-noise ratio using the coherent summation of pulses in the pulse train, the sample voltages are accumulated inside the Bedlam FPGAs. The accumulator buffer length is programmable with a buffer length of 8192 samples used in the experiments described here. A sampling rate of 1.024~GHz is used such that each buffer is 8~$\mu$sec long. The pulse generator is then triggered at a rate of 125~kHz so that one pulse is produced every 8~$\mu$sec. The trigger is synchronised with the pulse generator so the pulse voltages will be added coherently. The Bedlam board is programmed to dump the accumulated buffer using an Ethernet interface to a general purpose computer, when the required accumulation count is reached. At the pulse repetition rate of 125~kHz, and duration of 1~second, $1.25 \times 10^5$ pulses are accumulated resulting in a 350:1 improvement in the signal-to-noise ratio of the captured pulse. Longer integration periods can be used if desired. Initial tests showed some repetitive weak additive signals caused by sampling clock leakage into the ADC. Since these are clearly synchronous with the sampling clock they were removed by differencing the two halves of the output buffer with the pulse present in only one half, but the clock contamination equal in both halves. The Bedlam board functional schematic and accumulator block diagram are shown in Figs.~\ref{Bedlam_schem} and~\ref{Accum} respectively.

\begin{figure}[ht]
\center
\includegraphics[angle=0, width=0.9\linewidth]{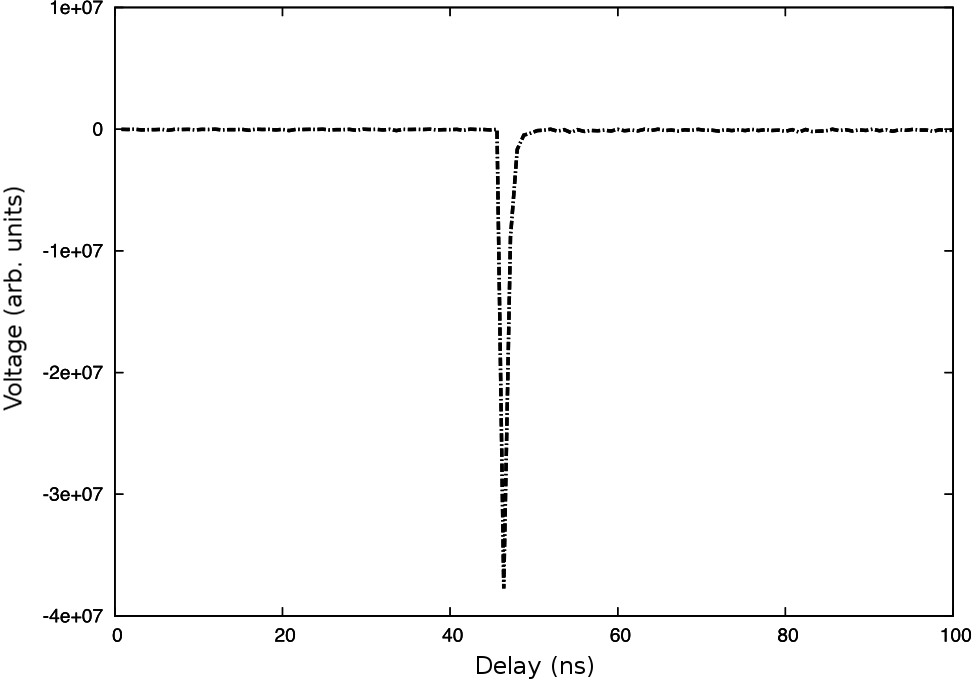}
\caption{Accumulator output after 1-second integration of the 1~MHz pulse train.}
\label{pulse_spect}
\end{figure}

Fig.~\ref{pulse_spect} shows a typical measured pulse, accumulated over 1~second, after differencing the two halves of the accumulation buffer. Only the first hundred samples are shown to show the output pulse profile. This is the pulse, fed directly to the accumulator, without the system being measured inserted in between the pulse generator and the accumulator. The Bedlam board has a voltage inversion at the input due to a balun that converts the single-ended to differential, so the pulse amplitude is negative at the output of the accumulator.  The polarity of the pulse, however, does not have a significant effect on the bandpass calibration.


\section{Experimental setup for bandpass measurements}
\label{sec:lab}

\begin{figure}[!htb]
\center
\includegraphics[ angle=0, width=\linewidth]{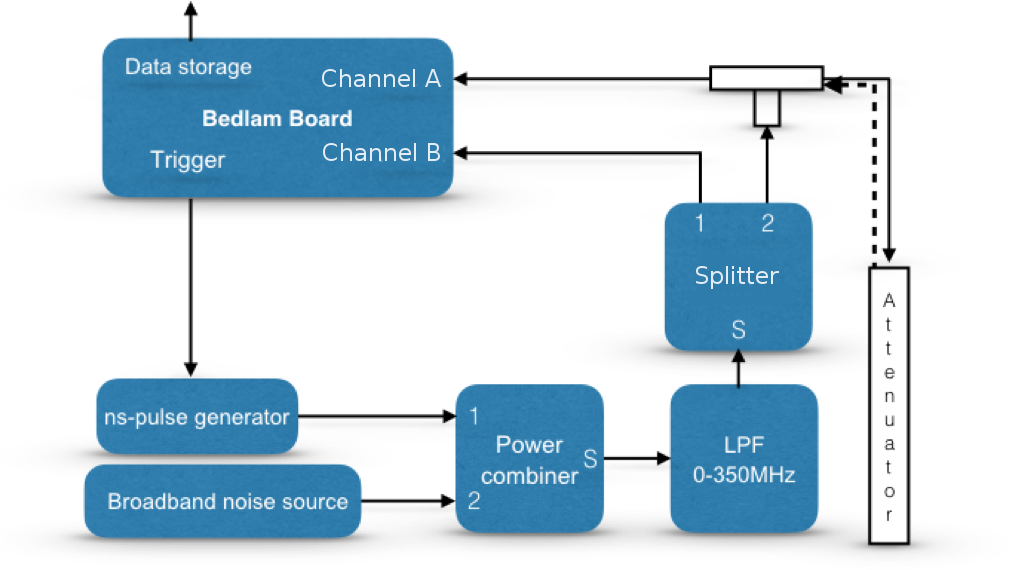}
\caption{Experimental setup for pulse calibration.  The power combiner, low-pass filter (LPF), splitter and T~connector together constitute the system under test.}
\label{SARAS_table_top}
\end{figure}

To validate our experimental setup for bandpass calibration, we performed a benchtop test of a system similar to the SARAS spectrometer~\cite{Patra13}.  This system contained a low-pass filter, and compared signal paths with and without a T~connector with an open load, to replicate the bandpass ripple observed in SARAS measurements.  The block diagram of the system under test is shown in Fig.~\ref{SARAS_table_top} with the pulse generator connected at its input and the accumulator connected to its output. Two channels of the Bedlam board are used with identical accumulator setup and common triggers. After injection, the short pulse is band limited by a lowpass filter with a cut off of 350~MHz. Filter output is split into two halves by a power splitter and one is fed directly to an accumulator via channel~B of the Bedlam board output which is used as a reference. The other half is fed to channel~A via a T~connector to mimic internal reflections that are typically present in receivers, and which must be calibrated. The third input of the T is connected to a 10~dB attenuator via a 5~m long cable and the output of the attenuator is connected to an open load. A part of the incident wave at the input of the T reached channel~A of the Bedlam board directly whereas the other part is reflected from the attenuator output and appears as a delayed pulse at the accumulator input in channel~A.

\begin{figure}[!htb]
\center
\includegraphics[angle=0, width=0.9\linewidth]{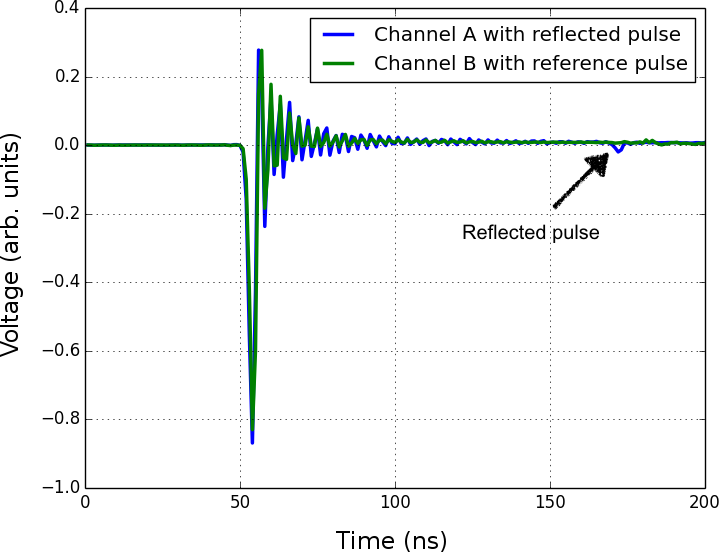}
\caption{Comparison of the channel~A and channel~B output after 1~second integration of the 1~MHz pulse train. Both channels contain the primary pulse which are identical and coincide in time. The pulse in channel A, however, goes through the T junction. A part of the pulse energy is lost due to the loss in the T junction and also a part of it is transmitted through the open end of the T which returns to channel A after reflection as a delayed pulse at a delay of 175~ns. Hence, the primary pulse amplitude in channel A is smaller compared to that in channel~B. }
\label{fig_7}
\end{figure}

Fig.~\ref{fig_7} shows the direct and the delayed pulses at the output of the accumulator in channel~A after 1~second integration.  Band limiting the signal to 350~MHz results in ringing of the pulse in the time domain. 
Fig.~\ref{fig_7} shows the reference pulse in channel~B and the pulse received in channel~A. Since the pulse power is divided between the direct and the reflected pulse, the direct pulse amplitude is smaller in channel~A compared to the same in channel~B.

%\begin{figure}[ht]
%\includegraphics[angle=0, height = 8.00cm, width=12.0cm]{Reflected_pulse_amp2.pdf}
%\caption{Channel~A output after 1 second integration of the 1~MHz pulse train. A part of the signal at the port 2 of the power splitter travels upto the open end of the bias-T and reflected back. The reflected pulse enters channel A and appears as the delayed pulse around 175 nS as shown in this figure.}
%\label{band_lim_spect}
%\end{figure}

\begin{figure}[!htb]
\center
\includegraphics[width=0.9\linewidth]{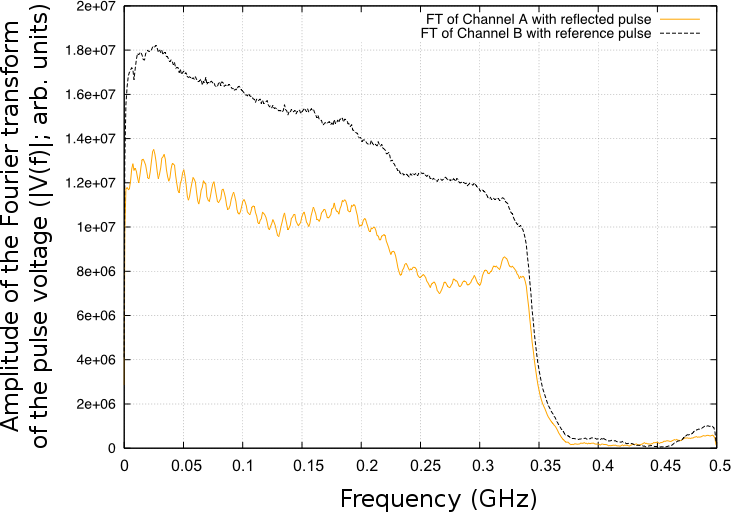}
\caption{Band limited output spectra. Y-axis is the Fourier transform of the voltage time samples.}
\label{band_lim_spect3}
\end{figure}

The true spectrum of the band limited pulse along channel~B with the spectrum of the pulse along channel~A are shown in Fig.~\ref{band_lim_spect3}.
Introduction of the T and long cable in channel~A have two obvious effects. In the time domain the delayed attenuated pulse is clearly seen with a delay of 120~ns and this caused the fast (16~MHz) ripple in the channel~A voltage amplitude spectrum. The voltage amplitude spectrum also shows a 320~MHz ripple which correspond to a 6~ns delay and is caused by reflection of the direct pulse between the Bedlam board input and the T output. The 6~ns delay is less conspicuous in the time domain as it is confused with the pulse oscillation caused by the sharp cut-off in the low pass filter.

%We have used this setup to make a number of tests of the stability of the coherent pulse calibration scheme. The sample variance for 600 samples each accumulated over 1~sec was $0.06\%$ for the direct pulse and the dynamic range in the 1~sec samples is 1500:1.  This system shows a DC drift over a time scale of 2--5~mins in the output at the same 1500:1 level which we have not pursued.

%Since the fractional power in the ns pulse is extremely small compared to the total system temperature, a pulse injection system can be run continuously during any observation to accurately determine the system bandpass response and calibrate the system output. The short duration of the generated pulse enables coherent accumulation of millions of pulses in just one second which results in an increase of 1000:1 in the signal-to-noise ratio compared to that of a of a single pulse.  This is a 60~dB increase in power. If the injected pulse power can be made  $>T_{\rm sys}$ in a system without saturation, the integration time could be substantially reduced.

As our results are consistent with the properties of the system under test, we consider this to be a successful benchtop validation of our bandpass calibration setup.  Assuming the input pulse to be sufficiently narrow (see section~\ref{sec:pulsegen}), Fig.~\ref{band_lim_spect3} then shows the amplitude of the system bandpass.

\section{Pulse calibration tests with the Parkes radio telescope}
\label{sec:parkes}

An opportunity arose to test the system described in section~\ref{sec:experiment} when three Bedlam boards were used at the Parkes radio telescope between 2--13 October (2013), in an experiment intended to detect pulsed radio emission from atmospheric particle cascades initiated by cosmic rays or gamma rays.  The Bedlam boards were originally designed for the LUNASKA experiment~\cite{LUNASKA}, in which an alternate mode was used to detect and store individual pulses coming from space.
 % The Bedlam boards were used in an alternate mode --- for which they were originally designed, for the LUNASKA experiment~\cite{LUNASKA} --- in which they detect and store individual pulses. 
 This experiment required high-precision timing calibration between the beams of the Parkes 21~cm multibeam receiver~\cite{Staveley_1996} in order to discriminate between radio-frequency interference, which typically appears in multiple beams simultaneously, and atmospheric particle cascades, which are expected to move from beam to beam with a timescale of nanoseconds.  This calibration was achieved with the same pulse generator system described in section~\ref{sec:experiment}, so we were able to carry out coherent pulse accumulation with the same hardware.

\begin{figure}[!htb]
\centering
\includegraphics[angle=0,width=10cm]{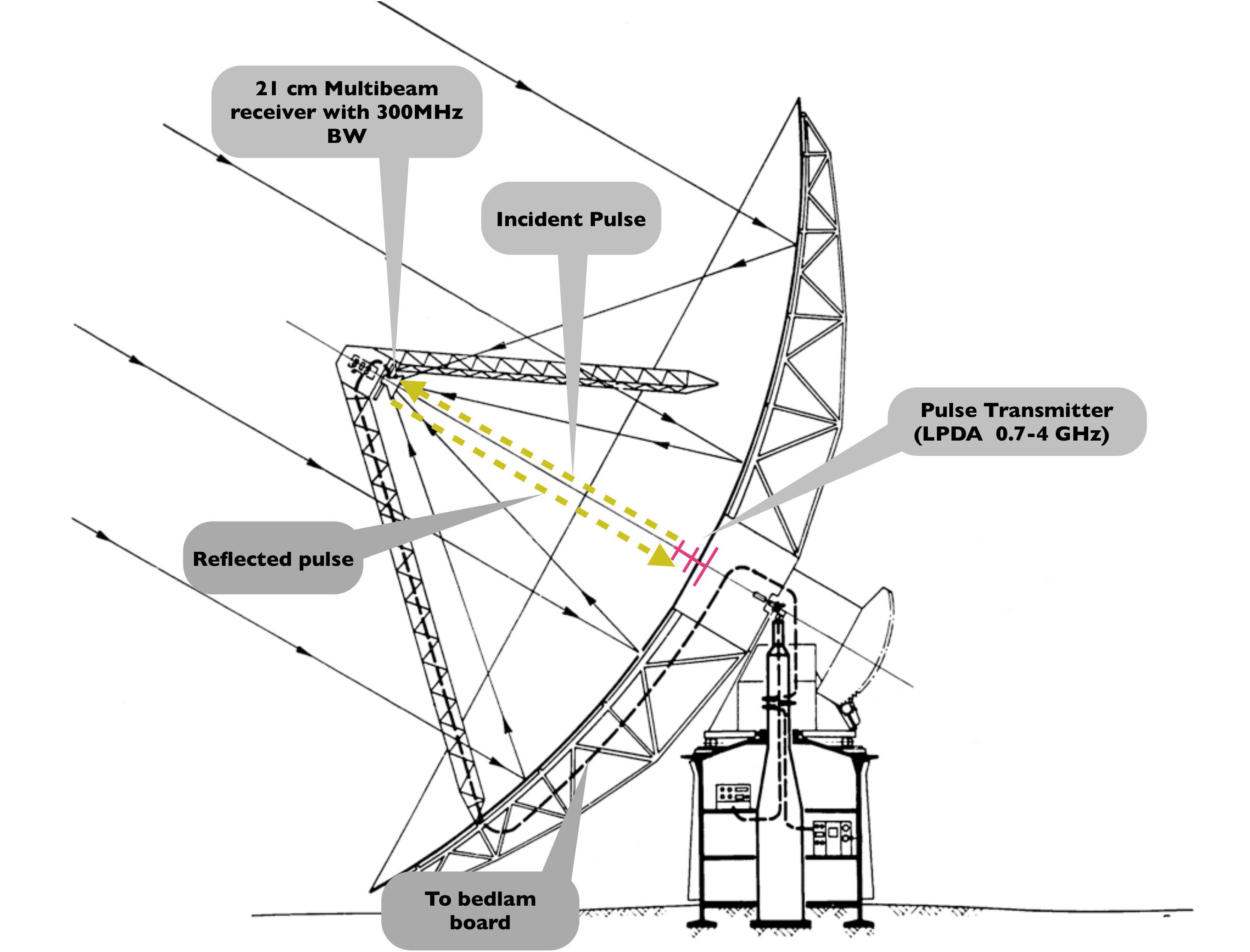}
\caption{Experimental setup for pulse calibration tests with the Parkes radio telescope.  The pulse is transmitted in the band 0.7--4~GHz from the (red) log-periodic dipole antenna (LPDA) at the vertex of the dish surface, and received by the 21~cm multibeam receiver at the focus with a narrower bandwidth (BW) of 300~MHz centred on 1.38~GHz.  Some fraction of the pulse is reflected (yellow) from the focus cabin, and is responsible for the reflected pulses described in the text.  The received signal is transmitted by cable (dashed) from the focus to a Bedlam board in the control tower.  (Schematic by J.M.~Sarkissian.)}
\label{Parkes_schem}
\end{figure}

The gross physical experimental setup is shown in Fig.~\ref{Parkes_schem}.  The signal from the pulse generator was transmitted from the apex of the parabolic dish, received at the focus, and ultimately returned to the Bedlam board, similar to Fig.~\ref{Accum}.  The system under test in this case comprises the large-scale structure of the telescope, the multibeam feed, and the remainder of the receiver signal path, including low-noise amplifiers, downconversion and attenuation.  Transmitting the pulse from the apex of the dish was convenient and flexible for this experiment, but may differ in some respects from observations of astronomical sources, which are in the far field of the telescope. 
%An alternate strategy in which we could the injected pulse into the feed for transmission back in to the dish would more accurately replicate an observation of radio source and this would be tried in future. 

\begin{figure}[ht]
\centering
\includegraphics[angle=0,width=0.9\linewidth]{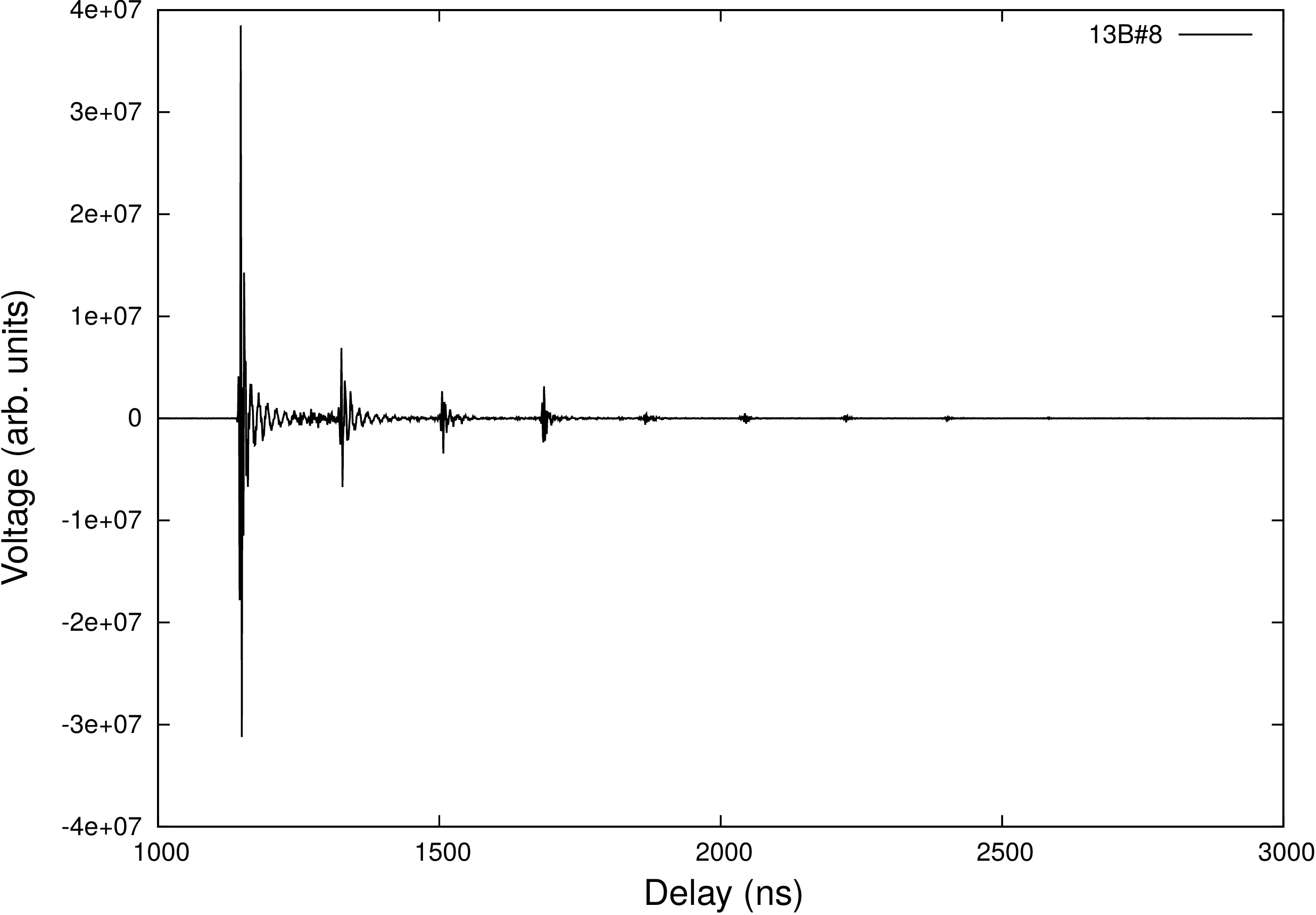}
\caption{Pulse profile recorded at Parkes using the pulse generator-accumulator system.  The spacing between the successively weaker pulses indicates that they result from repeated reflection of the pulse between the focus cabin and the vertex of the dish (see Fig.~\ref{Parkes_schem}).}
\label{Parkes_output}
\end{figure}

The pulse generator was set to produce pulses at a cadence of 125~kHz.  Fig.~\ref{Parkes_output} shows the accumulated pulse profile after one second of integration.  The direct pulse is most prominent, but there are eight additional pulses also visible, with successively smaller amplitudes, with a spacing of \mbox{$\sim 180$}~ns.  These result from the partial reflection of the transmitted pulse, first from the base of the focus cabin, then from the vertex of the dish.  These points are separated by 26~m, with a round-trip light travel time matching the observed pulse spacing.  In the frequency domain, these reflections create a ripple in the bandpass of the system, which is a well-known problem seen in most radio telescopes; for Parkes, the ripple has a period of 5.7~MHz~\cite{Barnes,Poulton}, matching the inverse of the pulse spacing.  The higher-order reflections (i.e.\ the second and subsequent reflected pulses) will result in harmonics which introduce sharper ripples in the system bandpass. 
A more detailed analysis of the reflected pulses reveals a complicated pattern of reflected polarized radiation which results from the structures around the telescope focus cabin.

\section{Conclusion}
\label{sec:conc}

We have explored a time domain technique for calibration of wideband spectrometers, which is particularly well suited for applications in radio astronomy.  This technique involves injecting a series of pulses into a system input, and coherently accumulating the system output, with no further need for complicated real-time processing: the amplitude and phase of the system bandpass response can be determined from the Fourier transform of this accumulated output.  The key advantages of this technique are that the total power contained in the individual pulses is negligible relative to the system noise, allowing it to run continuously without degrading system performance; and that it allows both the amplitude and phase of the bandpass response to be measured with a system implemented on simple hardware.  We have tested an implementation of this technique for bandpass amplitude calibration in a benchtop experiment, and with the 21~cm multibeam receiver system on the Parkes radio telescope.

The science goals of various proposed radio-astronomy experiments will require precise calibration of the system bandpass, which further development of our technique can in principle achieve.  For example, detection of the global signature of the epoch of reionization in the background radio spectrum requires a calibration accuracy of 1:$10^5$. Detection of baryon acoustic oscillations, spectral features imprinted on the redshifted 21~cm background at $z\sim1$, requires a calibration accuracy of 1:$10^6$.  Detection of the recombination of primordial plasma to form neutral hydrogen, which causes a broad spectral feature in the spectrum of the cosmic microwave background around 1--20~GHz, requires a calibration accuracy better than 1:$10^9$.  While this last case is beyond the time resolution possible with the current implementation of our technique, developments in fast sampling and signal processing may enable this in future experiments.

Our technique also has a potential application in the calibration of ultra-wideband receiver systems, which are currently being planned and constructed for several telescopes (e.g.~\cite{parkes_receiver}).  These receivers use a single feed with a bandwidth of, typically, several GHz, which is then split into several overlapping sub-bands before being sampled and digitised, to improve robustness against strong radio-frequency interference and to reduce the sampling-rate requirements.  After digitisation, these sub-bands are combined in order to restore the complete, continguous frequency band, but this requires calibration of the phase and amplitude differences between these sub-bands.  The coherency of our pulse calibration system allows this to be done, as the phase and amplitude of the pulse components must be continuous across the sub-band boundaries.

\begin{acknowledgements}
We would like to thank Raman Research Institute and Australia Telescope National Facility for providing the funding for this study. Parkes radio telescope is part of the Australia Telescope which is funded by the Commonwealth of Australia for operation as a National Facility managed by CSIRO.  JDB acknowledges support from ERC-StG 307215 (LODESTONE). NP acknowledges support from NSF CAREER award 1352519 and U.S. National Science Foundation (NSF) awards AST-1440343 and AST-1410719. The authors also declare that they have no conflict of interests.
\end{acknowledgements}


\begin{thebibliography}{}

\bibitem{Aubourg}
Aubourg, \'{E}ric et al.: Cosmological implications of baryon acoustic oscillation measurements.  Physical Review~D \textbf{92}, 123516 (2015)

\bibitem{Barnes}
Barnes, D.G., Briggs, F.H., Calabretta, M.R.: Post-correlation ripple removal and RFI rejection for Parkes Telescope survey data.  Radio Sci.\ \textbf{40}, RS5S13 (2005)

\bibitem{Bray13}
Bray, J.D., Ekers, R.D., Roberts, P.: Noise statistics in a fast digital radio receiver: the Bedlam backend for the Parkes Radio Telescope.  Exp.\ Astron.\ \textbf{36}, 155--174 (2013)

\bibitem{LUNASKA}
Bray, J.D., Ekers, R.D., Roberts, P., Reynolds, J.E., James, C.W., Phillips, C.J., Protheroe, R.J., McFadden, R.A., Aartsen, M.G.: A lunar radio experiment with the Parkes radio telescope for the LUNASKA project.  Astropart.\ Phys.\ \textbf{65}, 22--39 (2015)

%\bibitem{Hayes}
%Hayes, M.H.: Statistical digital signal processing and modeling.  John Wiley \& Sons, Hoboken, NJ (1996)

\bibitem{parkes_receiver}
Manchester, R.N. (for the PPTA team), Carretti, E., Norris, R.P., Phillips, C.J.: Development of an Ultra-Wideband (UWL) Receiver System at Parkes. Tech. Memo 40.3.2/002, Australia Telescope National Facility (2013)
 
\bibitem{Patra13}
Patra, N., Subrahmanyan, R., Raghunathan, A., Udaya Shankar, N.: SARAS: a precision system for measurement of the cosmic radio background and signatures from the epoch of reionization.  Exp.\ Astron.\ \textbf{36}, 319--370 (2013)

\bibitem{Poulton}
Poulton, G.T: Minimisation of Spectrometer Ripple In Prime Focus Radiotelescopes. CSIRO division of radio physics internal report (1974)

\bibitem{Shaver99}
 Shaver, P.A. et al.: Can the reionization epoch be detected as a global signature in the cosmic background?.  Astron.\ Astrophysics.\ \textbf{345}, 380--390 (1999)

\bibitem{Staveley_1996}
Staveley-Smith, L., Wilson, W.E., Bird, T.S., Disney, M.J., Ekers, R.D., Freeman, K.C., Haynes, R.F., Sinclair, M.W., Vaile, R.A., Webster, R.L., Wright, A.E.: The Parkes 21~cm multibeam receiver. PASA 13, 243--248 (Nov 1996)

\end{thebibliography}
\end{document}